# New Tolerance Factor to Predict the Stability of Perovskite Oxides and Halides

Christopher J. Bartel[1*], Christopher Sutton[2], Bryan R. Goldsmith[3], Runhai Ouyang[2], Charles B. Musgrave[1,4,5], Luca M. Ghiringhelli[2*], Matthias Scheffler[2]

[1]Department of Chemical and Biological Engineering, University of Colorado Boulder, Boulder, CO 80309, USA
[2]Fritz-Haber-Institut der Max-Planck-Gesellschaft, Faradayweg 4-6, D-14195 Berlin, Germany
[3]Department of Chemical Engineering, University of Michigan, Ann Arbor, MI 48109-2136, USA
[4]Department of Chemistry, University of Colorado Boulder, Boulder, CO 80309, USA
[5]Materials and Chemical Science and Technology Center, National Renewable Energy Laboratory, Golden, CO 80401, USA

**Abstract**

Predicting the stability of the perovskite structure remains a longstanding challenge for the discovery of new functional materials for many applications including photovoltaics and electrocatalysts. We developed an accurate, physically interpretable, and one-dimensional tolerance factor, $\tau$, that correctly predicts 92% of compounds as perovskite or nonperovskite for an experimental dataset of 576 $ABX_3$ materials ($X$ = $O^{2-}$, $F^-$, $Cl^-$, $Br^-$, $I^-$) using a novel data analytics approach based on SISSO (sure independence screening and sparsifying operator). $\tau$ is shown to generalize outside the training set for 1,034 experimentally realized single and double perovskites (91% accuracy) and is applied to identify 23,314 new double perovskites ($A_2BB'X_6$) ranked by their probability of being stable as perovskite. This work guides experimentalists and theorists towards which perovskites are most likely to be successfully synthesized and demonstrates an approach to descriptor identification that can be extended to arbitrary applications beyond perovskite stability predictions.

**Introduction**

Crystal structure prediction from chemical composition continues as a persistent challenge to accelerated materials discovery.(*1, 2*) Most approaches capable of addressing this challenge require several computationally demanding electronic-structure calculations for each material composition, limiting their use to a small set of materials.(*3-6*) Alternatively, descriptor-based approaches enable high-throughput screening applications because they provide rapid estimates of material properties.(*7, 8*) Notably, the Goldschmidt tolerance factor,(*9*) $t$, has been used extensively to predict the stability of the perovskite structure based only on the chemical formula, $ABX_3$, and the ionic radii, $r_i$, of each ion ($A$, $B$, $X$):

$$t = \frac{r_A + r_X}{\sqrt{2}(r_B + r_X)}. \qquad (1)$$

The perovskite crystal structure, as shown in **Fig. 1A**, is defined as any $ABX_3$ compound with a network of corner-sharing $BX_6$ octahedra surrounding a larger $A$-site cation ($r_A > r_B$), where the cations, $A$ and $B$, can span the periodic table and the anion, $X$, is typically a chalcogen or halogen. Distortions from the cubic structure can arise from size-mismatch of the cations and anion, which results in additional perovskite structures and nonperovskite structures. The *B* cation can also be replaced by two different ions, resulting in the double perovskite formula, $A_2BB'X_6$ (**Fig. 1B**). Single and double perovskite materials possess exceptional properties for a variety of applications such as electrocatalysis,(*10*) proton conduction,(*11*) ferroelectrics(*12*) (using oxides, $X = O^{2-}$), battery materials(*13*) (using fluorides, $X = F^-$), as well as photovoltaics(*14*) and optoelectronics(*15*) (using the heavier halides, $X = Cl^-, Br^-, I^-$).

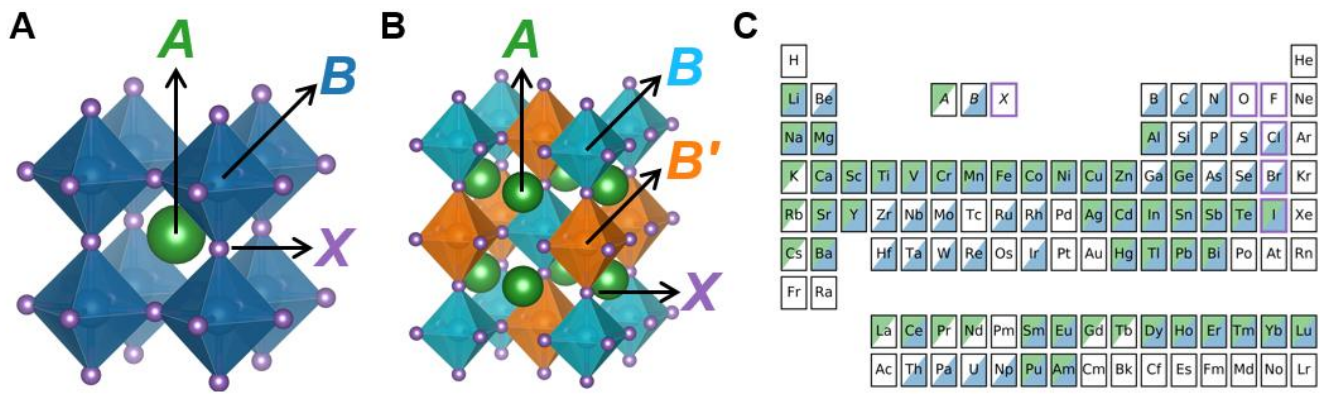

**Fig. 1. Perovskite structure and composition. A)** $ABX_3$, in the cubic single perovskite structure ($Pm\bar{3}m$), where the *A* cation is surrounded by a network of corner-sharing $BX_6$ octahedra. **B)** $A_2BB'X_6$, in the rock salt double perovskite structure ($Fm\bar{3}m$), where the *A* cations are surrounded by an alternating network of $BX_6$ and $B'X_6$ octahedra. In this structure, inverting the *B* and *B'* cations results in an equivalent structure. While the ideal cubic structures are shown here, perovskites may also adopt various non-cubic structures. **C)** A map of the elements that occupy the *A*, *B,* and/or *X* sites within the 576 compounds experimentally characterized as perovskite or nonperovskite at ambient conditions and reported in (*16-18*).

The first step in designing new perovskites for these applications is typically the assessment of stability using *t*, which has informed the design of perovskites for over 90 years. However, as reported in recent studies, its accuracy is often insufficient.(*19*) Considering 576 $ABX_3$ solids experimentally characterized at ambient conditions and reported in (*16-18*) (see **Fig. 1C** for the *A*, *B*, and *X* elements in this set), *t* correctly distinguishes between perovskite and nonperovskite for only 74% of materials and performs considerably worse for compounds containing heavier halides (chlorides – 51% accuracy, bromides – 56%, and iodides – 33%) than for oxides (83%) and fluorides (83%) (**Fig. 2A**, **Fig. S1**, **Table S1**). This deficiency in generalization to halide perovskites severely limits the applicability of *t* for materials discovery.

In this work, we present a new tolerance factor, *τ,* which has the form:

$$\tau = \frac{r_X}{r_B} - n_A \left( n_A - \frac{r_A/r_B}{\ln(r_A/r_B)} \right) \qquad (2)$$

where $n_A$ is the oxidation state of *A*, $r_i$ is the ionic radius of ion *i*, $r_A > r_B$ by definition and $\tau < 4.18$ indicates perovskite. A high overall accuracy of 92% for the experimental set (94% for a randomly chosen test set of 116 compounds) and nearly uniform performance across the five anions evaluated (oxides – 92% accuracy, fluorides – 92%, chlorides – 90%, bromides – 93%, iodides – 91%) is achieved with $\tau$ (**Fig. 2B**, **Fig. S1**, **Table S1**). Like *t*, the prediction of perovskite stability using $\tau$ requires only the chemical composition, allowing the tolerance factor to be agnostic to the many structures that are considered perovskite. In addition to predicting *if* a material is stable as perovskite, $\tau$ also provides a monotonic estimate of the *probability* that a material is stable in the perovskite structure. The accurate and probabilistic nature of $\tau$ as well as its generalizability over a broad range of single and double perovskites allows for new physical insights into the stability of the perovskite structure and the prediction of thousands of new double perovskite oxides and halides, 23,314 of which are provided here and ranked by their probability of being stable in the perovskite structure.

## Results and Discussion
### Finding an improved tolerance factor to predict perovskite stability

One key aspect of the performance of *t* is how well the sum of ionic radii estimates the interatomic bond distances for a given structure. Shannon's revised effective ionic radii,(*20*) based on a systematic empirical assessment of interatomic distances in nearly 1,000 compounds, are the typical choice for radii because they provide ionic radius as a function of ion, oxidation state, and coordination number for the majority of elements. Most efforts to improve *t* have focused on refining the input radii(*16, 18, 21, 22*) or increasing the dimensionality of the descriptor through two-dimensional structure maps(*17, 23, 24*) or high-dimensional machine learned models.(*25-27*) However, all hitherto applied approaches for improving the Goldschmidt tolerance factor are only effective over a limited range of $ABX_3$ compositions. Despite its modest classification accuracy, *t* remains the primary descriptor used by experimentalists and theorists to predict the stability of perovskites.

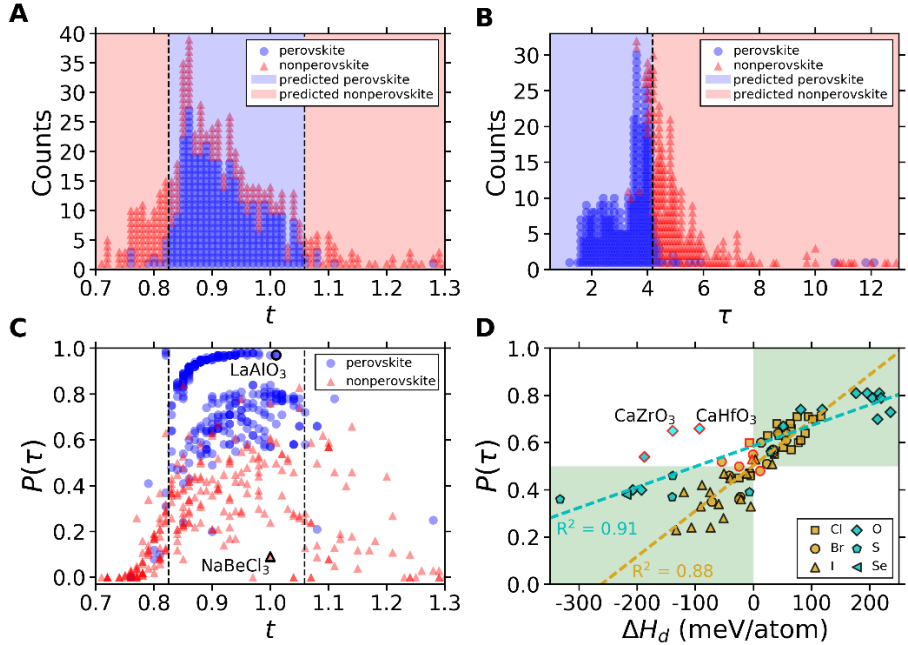

**Fig. 2**. **Assessing the performance of the improved tolerance factor, $\tau$. A)** A decision tree classifier determines that the optimal bounds for perovskite formability using the Goldschmidt tolerance factor, $t$, are $0.825 < t < 1.059$, which yields a classification accuracy of 74% for 576 experimentally characterized $ABX_3$ solids. **B)** $\tau$ achieves 92% classification accuracy on the set of 576 $ABX_3$ solids based on perovskite classification for $\tau < 4.18$, with this decision boundary identified using a one-node decision tree. All classifications made by $t$ and $\tau$ on the experimental dataset are provided in **Table S1**. The largest value of $\tau$ in the experimental set of 576 compounds is 181.5, however, all points with $\tau > 13$ are correctly labeled as nonperovskite and not shown to highlight the decision boundary. The outlying compounds at $\tau > 10$ that are labeled perovskite yet have large $\tau$ are $PuVO_3$, $AmVO_3$, and $PuCrO_3$, which may indicate poorly defined radii or incorrect experimental characterization. **C)** A comparison of Platt-scaled classification probabilities, $P(\tau)$, versus $t$. $LaAlO_3$ and $NaBeCl_3$ are labeled to highlight the variation in $P(\tau)$ at nearly constant $t$. **D)** A comparison between $P(\tau)$ and the decomposition enthalpy ($\Delta H_d$) for 36 double perovskite halides calculated using density functional theory (DFT) in the $Fm\overline{3}m$ structure in (*28*), and 37 single and double perovskite chalcogenides and halides in the $Pm\overline{3}m$ structure in (*29*). The legend corresponds with the anion, $X$. Positive decomposition enthalpy ($\Delta H_d > 0$) indicates the structure is stable with respect to decomposition into competing compounds. The green and white shaded regions correspond with agreement and disagreement between the calculated $\Delta H_d$ and the classification by $\tau$. Points of disagreement are outlined in red. $CaZrO_3$ and $CaHfO_3$ are labeled because they are known to be stable in the perovskite structure, although they are unstable in the cubic structure.(*30, 31*) For this reason, the best fit line for the chalcogenides ($X = O^{2-}$, $S^{2-}$, $Se^{2-}$) excludes these two points.

The SISSO (sure independence screening and sparsifying operator) approach(*32*) was used to identify an improved tolerance factor for predicting whether a given compound is perovskite (determined by experimental realization of any structure with corner-sharing $BX_6$ octahedra (*21*) at ambient conditions) or nonperovskite (determined by experimental realization of any structure(s) without corner-sharing $BX_6$ octahedra, including in some cases, failed synthesis of any $ABX_3$ compound). Of the 576 experimentally characterized $ABX_3$ solids, 80% were used to train and 20% to test the SISSO-learned descriptor. Several alternative atomic properties were considered as candidate features and, among them, SISSO determined that the best performing descriptor, $\tau$ (**Equation 2**, **Fig. 2B**), depends only on oxidation states and Shannon ionic radii. See **Methods** for an explanation of the approach used for descriptor identification and a discussion of alternative approaches. For the set of 576 $ABX_3$ compositions, $\tau$ correctly labels 94% of the perovskites and 89% of the nonperovskites compared with 94% and 49%, respectively, using $t$. The primary advantage of $\tau$ over $t$ is the remarkable reduction in false positives – compounds predicted to be perovskite but are not experimentally identified as stable perovskites – with false positive rates for $\tau$ and $t$ of 11% and 51%, respectively. Full confusion matrices along with additional performance metrics for $\tau$ and $t$ are provided in **Table S2**. The large decrease in false positive rate (from 51% to 11%) while substantially increasing the overall classification accuracy (from 74% to 92%) demonstrates that $\tau$ improves significantly upon $t$ as a reliable tool to guide experimentalists towards which compounds can be synthesized in perovskite structures.

Beyond the improved accuracy, a significant advantage of $\tau$ is the monotonic (continuous) dependence of perovskite stability on $\tau$. As $\tau$ decreases, the $\tau$-based probability of being perovskite, $P(\tau)$, increases, where perovskites are expected for an empirically determined range of $\tau < 4.18$ (**Fig. 2B**, **Methods** for details). Probabilities are obtained using Platt's scaling,(*33*) where the *binary classification* of perovskite/nonperovskite is transformed into a *continuous probability* estimate of perovskite stability, $P(\tau)$, by training a logistic regression model on the $\tau$-derived binary classification. Probabilities cannot similarly be obtained with $t$ because the stability of the perovskite structure does not increase or decrease monotonically with $t$, where $0.825 < t < 1.059$ results in a classification as perovskite (this range maximizes the classification accuracy of $t$ on the set of 576 compounds). While $P(\tau)$ is sigmoidal with respect to $\tau$ because of the logistic fit (**Fig. S2**), a bell-shaped behavior of $P(\tau)$ with respect to $t$ is observed because of the multiple decision boundaries required for $t$ (**Fig. 2C**). This relationship leads to an increase in $P(\tau)$ (i.e., probability of perovskite stability using $\tau$) with an increase in $t$ until a value of $t \sim 0.9$. Beyond this range, the probabilities level out or decrease as $t$ increases further.

The disparity between the $\tau$-derived perovskite probability, $P(\tau)$, and the assignment by $t$ can be significant, especially in the range where $t$ predicts a stable perovskite ($0.825 < t < 1.059$). A comparison of the perovskite, $LaAlO_3$, and the nonperovskite, $NaBeCl_3$, illustrates the discrepancy between these two approaches. $t$ incorrectly predicts both compounds to be perovskite ($t = 1.0$) whereas $P(\tau)$ varies from $< 10\%$ for $NaBeCl_3$ to $> 97\%$ for $LaAlO_3$, in agreement with experiment. For $NaBeCl_3$, instability in the perovskite structure arises from an insufficiently large $Be^{2+}$ cation on the *B*-site, which leads to unstable $BeCl_6$ octahedra. This contribution to perovskite stability is accounted for in the first term of $\tau$ (**Equation 2**, $r_X/r_B = \mu^{-1}$, where $\mu$ is the octahedral factor).

$\mu$ is the typical choice for a second feature used in combination with $t$,(*17, 18, 23*) and was recently used to assess the predictive accuracy of Goldschmidt's "no-rattling" principle. In this analysis, six inequalities dependent upon $t$ and $\mu$ were derived and used to predict the formability of single and double perovskites with a reported accuracy of ~80%.(*34*) Notably, training a decision tree algorithm on the bounds of $t$ and $\mu$ that optimally separate perovskite from nonperovskite leads to an 85% classification accuracy for this dataset (**Fig. S3**). In contrast to these two-dimensional descriptors based on ($t, \mu$), $\tau$ incorporates $\mu$ as a one-dimensional descriptor yet still achieves a higher accuracy of 92%, demonstrating the capability of the SISSO algorithm to identify a highly accurate tolerance factor comprised of intuitively meaningful parameters.

The nature of geometrical descriptors, such as $t$ or $\mu$, are fundamentally different than data-driven descriptors, such as $\tau$. $t$ and $\mu$ are derived from geometric constraints that indicate when the perovskite structure is a possible structure that can form. However, these constraints do not necessarily indicate when the perovskite structure is the ground-state structure and does, in fact, form. For instance, if $t = 1$ and the ionic limit on which $t$ was derived is applicable (the interatomic distances are sums of the ionic radii), these criteria do not suggest perovskite is the ground-state structure, only that the interatomic distances are such that the lattice constants in the $A$-$X$ and $B$-$X$ directions can be commensurate with the perovskite structure. The fact that $t$ does not guarantee the formation of the perovskite structure is evident by the high false positive rate (51%) in the region of $t$ where perovskite is expected ($0.825 < t < 1.059$). Similarly, although $\mu$ may fall within the range where $BX_6$ octahedra are expected based on geometric considerations ($0.414 < \mu < 0.732$), the octahedra that form may be edge- or face-sharing and therefore the observed structure is nonperovskite. In this work, SISSO searches a massive space of potential descriptors to identify the one that most successfully detects when a given chemical formula will or will not crystallize in the perovskite structure and because this is the target property, $\tau$ emerges as a much more predictive descriptor than $t$ or $\mu$.

Although the classification by $\tau$ disagrees with the experimental label for 8% of the 576 compounds, the agreement increases to 99% outside the range $3.31 < \tau < 5.92$ (200 compounds) and 100% outside the range $3.31 < \tau < 12.08$ (152 compounds). The experimental dataset may also be imperfect as compounds can manifest different crystal structures as a function of the synthesis conditions due to, for example, defects in the experimental samples (impurities, vacancies, etc.). These considerations emphasize the usefulness of $\tau$-derived probabilities, in addition to the binary classification of perovskite/nonperovskite, which address these uncertainties in the experimental data and corresponding classification by $\tau$.

**Comparing $\tau$ to calculated perovskite stabilities**

The precise and probabilistic nature of $\tau$, as well as its simple functional form – depending only on widely available Shannon radii (and the oxidation states required to determine the radii) – enables the rapid search across composition space for stable perovskite materials. Prior to attempting synthesis, it is common for new materials to be examined using computational approaches, therefore it is useful to compare the predictions from $\tau$ with those obtained using density functional theory. The stabilities (decomposition enthalpies, $\Delta H_d$) of 73 single and double perovskite chalcogenides and halides were recently examined with density functional theory utilizing the Perdew-Burke-Ernzerhof(*35*) exchange-correlation functional (DFT).(*28, 29*) $\tau$ is found to agree with the calculated stability for 64 of 73 calculated materials. Significantly, the probabilities that result from classification with $\tau$ linearly correlate with $\Delta H_d$, demonstrating the value of the monotonic behavior of $\tau$ and $P(\tau)$ (**Fig. 2D**, **Table S3**).

Although $\tau$ appears to disagree with these DFT calculations for nine compounds, six disagreements lie near the decision boundaries ($P(\tau) = 0.5$, $\Delta H_d = 0$ meV/atom), suggesting that they cannot be confidently classified as stable or unstable perovskites using $\tau$ or DFT calculations of the cubic structure. Of the remaining disagreements, CaZrO$_3$ and CaHfO$_3$ reveal the power of $\tau$ compared with DFT calculations of the cubic structure as these two oxides are known to be isostructural with the orthorhombic perovskite CaTiO$_3$, from which the name perovskite originates.(*30, 31*) $\Delta H_d <$ −90 meV/atom for these two compounds in the cubic structure, indicating they are nonperovskites. In contrast, $\tau$ predicts both compounds to be stable perovskites with ~65% probability, which agrees with experiment. These results show that a key challenge in the prediction of perovskite stability from quantum chemical calculations is the requirement of a specific structure as an input as there are more than a dozen unique structures classified as perovskite (i.e., those having corner-sharing $BX_6$ octahedra) and many more which are nonperovskite.

Several recent machine-learned descriptors for perovskite stability have been trained or tested on DFT-calculated stabilities of only the cubic perovskite structure.(*29, 36-38*) However, less than 10% of perovskites are observed experimentally in this structure,(*21*) leading to an inherent disagreement between the descriptor predictions and experimental observations. Recently, it was shown that of 254 synthesized perovskite oxides (*ABO$_3$*), DFT calculations in the Open Quantum Materials Database (OQMD) (*39*) predict only 186 (70%) to be stable or even moderately unstable (within 100 meV/atom of the convex hull).(*27*) The discrepancy is likely associated with the difference in energy between the true perovskite ground state and the calculated high-symmetry structure(s). Because $\tau$ was trained exclusively on the experimental characterization of $ABX_3$ compounds, $\tau$ is informed by the true ground state (or metastable but observed) structure of each $ABX_3$ and the potential for these compounds to decompose into any compound(s) in the *A-B-X* composition space. A principal advantage of $\tau$ over many existing descriptors is that its identification and validation were based on experimentally observed stability or instability of a structurally diverse dataset.

**Extension to double perovskite oxides and halides**

Double perovskites are particularly intriguing as an emerging class of semiconductors that offer a lead-free alternative to traditional perovskite photoabsorbers and increased compositional tunability for enhancing desired properties such as catalytic activity.(*10, 19, 40*) Still, the experimentally realized composition space of double perovskites is relatively unexplored compared with the number of possible *A*, *B*, *B'*, and *X* combinations that can form $A_2BB'X_6$ compounds. The set of 576 compounds used for training and testing $\tau$ are comprised of 49 *A* cations, 67 *B* cations, and 5 *X* anions, from which > 500,000 double perovskite formulas, $A_2BB'X_6$, can be constructed. Comparing with the Inorganic Crystal Structure Database (ICSD)(*34, 41*) reveals only 918 compounds (< 0.2%) with known crystal structures, 868 of which are perovskite.

Although $\tau$ was only trained on $ABX_3$ compounds, it is readily adaptable to double perovskites because it depends only on composition and not structure. To extend $\tau$ to $A_2BB'X_6$ formulas, $r_B$ is approximated as the arithmetic mean of the two B-site radii ($r_B$, $r_{B'}$). $\tau$ correctly classifies 91% of these 918 $A_2BB'X_6$ compounds in the ICSD (compared with 92% on 576 $ABX_3$ compounds), recovering 806 of 868 known double perovskites (**Table S4**). The geometric mean has also been used to approximate the radius of a site with two ions.(*42*) We find this has little effect on classification with $\tau$ as 91% of the 918 $A_2BB'X_6$ compounds are also correctly classified using the geometric mean for $r_B$, and the classification label differs for only 14 of 918 compounds using the arithmetic or geometric mean. Although $\tau$ was identified using 460 $ABX_3$ compounds, the agreement with experiment on these compounds (92%) is comparable to that on the 1,034 compounds (91%) that span $ABX_3$ (116 compounds) and $A_2BB'X_6$ (918 compounds) formulas and were completely excluded from the development of $\tau$ (i.e., test set compounds). This result indicates significant generalizability to predicting experimental realization for single and double perovskites that are yet to be discovered. With $\tau$ thoroughly validated as being predictive of experimental stability, the space of yet-undiscovered double perovskites was explored to identify 23,314 charge-balanced double perovskites that $\tau$ predicts to be stable at ambient conditions (of > 500,000 candidates). These compounds are provided in **Table S4** including assigned oxidation states and radii along with $t$, $\tau$, predictions made using each tolerance factor, and classification in the ICSD where available. Importantly, there are many thousands of additional compounds with substitutions on the A and/or X sites – $AA'BB'(XX')_6$ – that are expected to be similarly rich in yet-undiscovered perovskite compounds.

Two particularly attractive classes of materials within this set of $A_2BB'X_6$ compounds are double perovskites with $A = Cs^+$, $X = Cl^-$ and $A = La^{3+}$, $X = O^{2-}$ which have garnered significant interest in a number of applications including photovoltaics, electrocatalysis, and ferroelectricity. The ICSD contains 45 compounds (42 perovskites) with the formula $CsBB'Cl_6$, 43 of which are correctly classified as perovskite or nonperovskite by $\tau$. From the high-throughput analysis using $\tau$, we predict an additional 420 perovskites to be stable with 164 having at least the probability of perovskite formation as the recently synthesized perovskite, $Cs_2AgBiCl_6$ ($P(\tau) = 69.6\%$).(*43*) A map of perovskite probabilities for charge-balanced $Cs_2BB'Cl_6$ compounds is shown in **Fig. 3** (lower triangle). Within this set of 164 probable perovskites, there is significant opportunity to synthesize double perovskite chlorides that contain $3d$ transition metals substituted on one or both B-sites as 83 new compounds of this type are predicted to be stable as perovskite with high probability.

While double perovskite oxides have been explored extensively for a number of applications, the small radius and favorable charge of $O^{2-}$ yields a massive design space for the discovery of new compounds. For $La_2BB'O_6$, ~63% of candidate compositions are found to be charge-balanced compared with only ~24% of candidate $Cs_2BB'Cl_6$ compounds. The ICSD contains 85 $La_2BB'O_6$ compounds, *all of which* are predicted to be perovskite by $\tau$ in agreement with experiment. We predict an additional 1,128 perovskites to be discoverable in this space, with a remarkable 990 having $P(\tau) \geq 85\%$ (**Fig. 3**, upper triangle). All 128 $ABX_3$ compounds in the experimental set that meet this threshold are experimentally realized as perovskite, suggesting there is ample opportunity for perovskite discovery in lanthanum oxides.

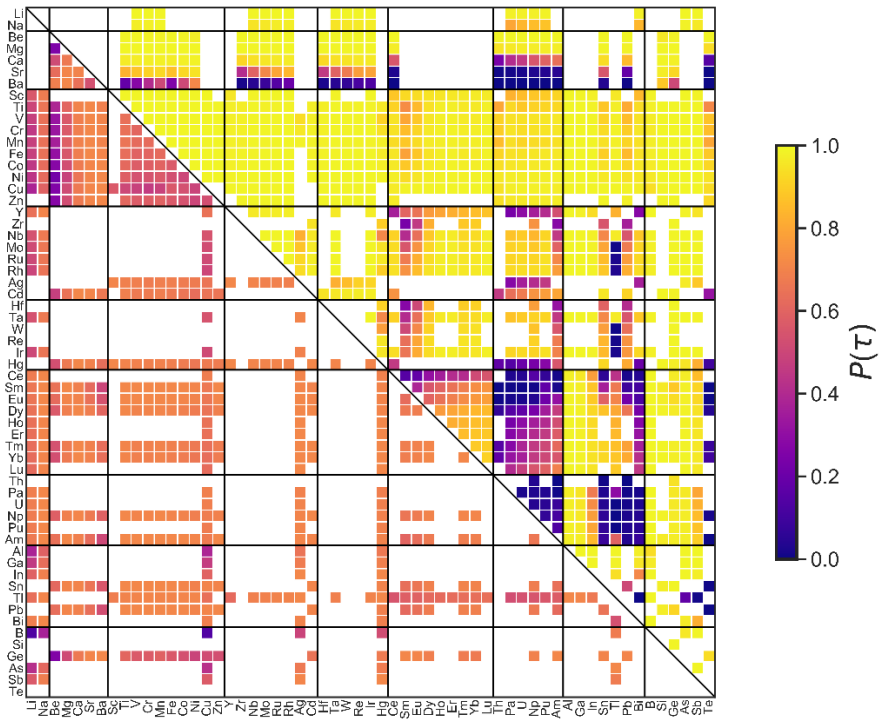

**Fig. 3. Map of predicted double perovskite oxides and halides.** *Lower triangle*: the probability of forming a stable perovskite with the formula $Cs_2BB'Cl_6$ as predicted by $\tau$. *Upper triangle*: the probability of forming a stable perovskite with the formula $La_2BB'O_6$ as predicted by $\tau$. White spaces indicate $B/B'$ combinations that do not result in charge-balanced compounds with $r_A > r_B$. The colors indicate the Platt-scaled classification probabilities, $P(\tau)$, with higher $P(\tau)$ indicating a higher probability of forming a stable perovskite. $B/B'$ sites are restricted to ions that are labeled as $B$ sites in the experimental set of 576 $ABX_3$ compounds.

**Compositional mapping of perovskite stability**

In addition to enabling the rapid exploration of stoichiometric perovskite compositions, $\tau$ provides the probability of perovskite stability, $P(\tau)$, for an arbitrary combination of $n_A$, $r_A$, $r_B$, and $r_X$, which is shown in **Fig. 4**. For each grouping shown in **Fig. 4**, experimentally realized perovskites and nonperovskites are shown as single points to compare with the range of values in the predictions made from $\tau$. Doping at various concentrations presents a nearly infinite number of $A_{1-x}A'_xB_{1-y}B'_y(X_{1-z}X'_z)_3$ compositions that allows for the tuning of technologically useful properties. $\tau$ suggests the size and concentration of dopants on the $A$, $B$, or $X$ sites that likely lead to improved stability in the perovskite structure. Conversely, compounds that lie in the high-probability region are likely amenable to ionic substitutions that decrease the probability of forming a perovskite, but may improve a desired property for another application. For example, $LaCoO_3$, with $P(\tau) = 98.9\%$, should accommodate reasonable ionic substitutions (i.e., $A$-sites of comparable size to La or $B$-sites of comparable size to Co), and was recently shown to have enhanced oxygen exchange capacity and nitric oxide oxidation kinetics with stable substitutions of Sr on the *A*-site.(*44*)

The probability maps in **Fig. 4** arise from the functional form of $\tau$ (**Equation 2**) and provide insights into the stability of the perovskite structure as the size of each ion is varied. The perovskite structure requires that the *A* and *B* cations occupy distinct sites in the $ABX_3$ lattice, with *A* 12-fold and *B* 6-fold coordinated by *X*. When $r_A$ and $r_B$ are too similar, nonperovskite lattices that have similarly coordinated *A* and *B* sites, such as cubic bixbyite, become preferred over the perovskite structure. Based on the construct of $\tau$, as $r_A/r_B \to 1$, $P(\tau) \to 0$, which arises from the $+x/\ln(x)$ ($x = r_A/r_B$) term, where $\lim_{x \to 1} \frac{x}{\ln(x)} = +\infty$ and larger values of $\tau$ lead to lower probabilities of forming perovskites. When $r_A = r_B$, $\tau$ is undefined, yet compounds where *A* and *B* have identical radii are rare and not expected to adopt perovskite structures ($t = 0.71$).

The octahedral term in $\tau$ ($r_X/r_B$) also manifests itself in the probability maps, particularly in the lower bound on $r_B$ where perovskites are expected as $r_X$ is varied. As $r_X$ increases, $r_B$ must similarly increase to enable the formation of stable $BX_6$ octahedra. This effect is noticeable when separately comparing compounds containing $Cl^-$ (left), $Br^-$ (center), and $I^-$ (right) (bottom row of **Fig. 4**), where the range of allowed cation radii decreases as the anion radius increases. For $r_B \ll r_X$, $r_X/r_B$ becomes large, which increases $\tau$ and therefore decreases the probability of stability in the perovskite structure. This accounts for the inability of small *B*-site ions to sufficiently separate *X* anions in $BX_6$ octahedra, where geometric arguments suggest that *B* is sufficiently large to form $BX_6$ octahedra only for $r_B/r_X > 0.414$. Because the cation radii ratios significantly affect the probability of perovskite, as discussed in the context of $x/\ln(x)$, $r_X$ also has a significant indirect effect on the lower bound of $r_A$, which increases as $r_X$ increases.

The role of $n_A$ in $\tau$ is more difficult to parse, but its placement dictates two effects on stability – as *A* is more oxidized (increasing $n_A$), $-n_A^2$ increases the probability of forming the perovskite structure, but $n_A$ also magnifies the effect of the $x/\ln(x)$ term, increasing the importance of the cation radii ratio. Notably, $n_A = 1$ for most halides and some oxides (245 of the 576 compounds in our set) and in these cases, $\tau = \frac{r_X}{r_B} + \frac{r_A/r_B}{\ln(r_A/r_B)} - 1$ for all combinations of *A, B,* and *X* and $n_A$ plays no role as the composition is varied.

This analysis illustrates how data-driven approaches can be used to not only maximize the predictive accuracy of new descriptors, but can also be leveraged to understand the actuating mechanisms of a target property – in this case, perovskite stability. This attribute distinguishes $\tau$ from other descriptors for perovskite stability that have emerged in recent years. For instance, three recent works have shown that the experimental formability of perovskite oxides and halides can be separately predicted with high accuracy using kernel support vector machines,(*26*) gradient boosted decision trees,(*25*) or a random forest of decision trees.(*27*) While these approaches can yield highly accurate models, the resulting descriptors are not documented analytically, and therefore, the mechanism by which they make the perovskite/nonperovskite classification is opaque.

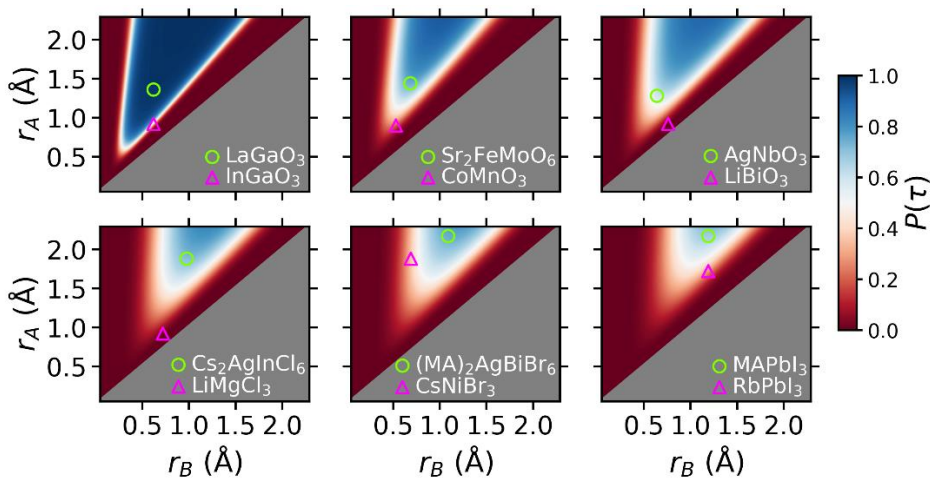

**Fig. 4. The effects of ionic radii and oxidation states on the stability of single and double perovskite oxides and halides.** *Top row*: $X = O^{2-}$ (left to right – $n_A = 3^+, 2^+, 1^+$). *Bottom row*: $n_A = 1^+$ (left to right – $X = Cl^-, Br^-, I^-$). The experimentally realized perovskites $LaGaO_3$, $Sr_2FeMoO_6$, $AgNbO_3$, $Cs_2AgInCl_6$, $(MA)_2AgBiBr_6$, and $MAPbI_3$ are shown as open circles in the corresponding plot, which are all predicted to be stable by $\tau$. The experimentally realized nonperovskites $InGaO_3$, $CoMnO_3$, $LiBiO_3$, $LiMgCl_3$, $CsNiBr_3$, and $RbPbI_3$ are shown as open triangles and predicted to be unstable in the perovskite structure by $\tau$. The organic molecule, methylammonium (MA), is shown in the last two panels. While $(MA)_2AgBiBr_6$ and $MAPbI_3$ are correctly classified with $\tau$, only inorganic cations were used for descriptor identification; therefore, $r_A = 1.88$ Å ($Cs^+$) is the largest cation considered. The gray region where $r_B > r_A$ is not classified because when this occurs, $A$ becomes $B$ and vice versa based on our selection rule $r_A > r_B$.

**Conclusions**

We report a new tolerance factor, $\tau$, that enables the prediction of experimentally observed perovskite stability significantly better than the widely used Goldschmidt tolerance factor, $t$, and the two-dimensional structure map using $t$ and the octahedral factor, $\mu$. For 576 $ABX_3$ and 918 $A_2BB'X_6$ compounds, the prediction by $\tau$ agrees with the experimentally observed stability for > 90% of compounds, with > 1,000 of these compounds reserved for testing generalizability (prediction accuracy). The deficiency of $t$ arises from its functional form and not the input features as the calculation of $\tau$ requires the same inputs as $t$ (composition, oxidation states, and Shannon ionic radii). Thus, $\tau$ enables a superior prediction of perovskite stability with negligible computational cost. The monotonic and one-dimensional nature of $\tau$ allows for the determination of perovskite probability as a continuous function of the radii and oxidation states of $A$, $B$, and $X$. These probabilities are shown to linearly correlate with DFT-computed decomposition enthalpies and help clarify how chemical substitutions at each of the sites modulate the tendency for perovskite formation. Using $\tau$, we predict the probability of double perovskite formation for thousands of unexplored compounds, resulting in a library of stable perovskites ordered by their likelihood of forming perovskites. Due to the simplicity and accuracy of $\tau$, we expect its use to accelerate the discovery and design of state-of-the-art perovskite materials for applications ranging from photovoltaics to electrocatalysis.

## Materials and Methods

### Radii assignment

To develop a descriptor that takes as input the chemical composition and outputs a prediction of perovskite stability, the features that comprise the descriptor must also be based only on composition. Yet it is not known *a priori* which cation will occupy the *A*-site or *B*-site given only a chemical composition, $CC'X_3$ (*C* and *C'* being cations). To determine which cation is *A* or *B*, a list of allowed oxidation states (based on Shannon's radii(*20*)) is defined for each cation. All pairs of oxidation states for *C* and *C'* that charge-balance $X_3$ are considered. If more than one charge-balanced pair exists, a single pair is chosen based on the electronegativity ratio of the two cations ($\chi_C/\chi_{C'}$). If $0.9 < \chi_C/\chi_{C'} < 1.1$, the pair that minimizes $|n_C - n_{C'}|$ is chosen, where $n_C$ is the oxidation state for *C*. Otherwise, the pair that maximizes $|n_C - n_{C'}|$ is chosen. With the oxidation states of *C* and *C'* assigned, the values of the Shannon radii for the cations occupying the *A* and *B* sites are chosen to be closest to the coordination number of twelve and six, which are consistent with the coordination environments of the *A* and *B* cations in the perovskite structure. Finally, the radii of the *C* and *C'* cations are compared and the larger cation is assigned as the *A*–site cation. This strategy reproduces the assignment of the *A* and *B* cations for 100% of 313 experimentally labeled perovskites.

### Selection of $\tau$

For the selection of $\tau$, the oxidation states ($n_A$, $n_B$, $n_X$), ionic radii ($r_A$, $r_B$, $r_X$), and radii ratios ($r_A/r_B$, $r_A/r_X$, $r_B/r_X$) comprise the primary features, $\Phi_0$, where $\Phi_n$ refers to the descriptor-space with *n* iterations of complexity as defined in (*32*). For example, $\Phi_1$ refers to the primary features ($\Phi_0$) together with one iteration of algebraic/functional operations applied to each feature in $\Phi_0$. $\Phi_2$ then refers to the application of algebraic/functional operations to all potential descriptors in $\Phi_1$, and so forth. Note that $\Phi_m$ contains all potential descriptors within $\Phi_{n<m}$ with a filter to remove redundant potential descriptors. For the discovery of $\tau$, complexity up to $\Phi_3$ is considered, yielding $\sim 3 \times 10^9$ potential descriptors. An alternative would be to exclude the radii ratios from $\Phi_0$ and construct potential descriptors with complexity up to $\Phi_4$. However, given the minimal $\Phi_0 = [n_A, n_B, n_X, r_A, r_B, r_X]$, there are $\sim 10^8$ potential descriptors in $\Phi_3$, so $\sim 10^{16}$ potential descriptors would be expected in $\Phi_4$ (based on $\sim 10^2$ being present in $\Phi_1$ and $\sim 1 \times 10^4$ in $\Phi_2$), and this number is impractical to screen using available computing resources.

The dataset of 576 $ABX_3$ compositions was partitioned randomly into an 80% training set for identifying candidate descriptors and a 20% test set for analyzing the predictive ability of each descriptor. The top 100,000 potential descriptors most applicable to the perovskite classification problem were identified using one iteration of SISSO with a subspace size of 100,000. Each descriptor in the set of $\sim 3 \times 10^9$ was ranked according to domain overlap, as described in Ouyang *et al.*(*32*) To identify a decision boundary for classification, a decision tree classifier with a max depth of two was fit to the top 100,000 candidate descriptors ranked based on domain overlap. Domain overlap (and not decision tree performance) is used as the SISSO ranking metric because of the significantly lower computational expense associated with applying this metric. Notably, $\tau$ was the 14,467$^{th}$ highest ranked descriptor by SISSO using the domain overlap metric and, as such, this defines the minimum subspace required to identify $\tau$ using this approach. Without evaluating a decision tree model for each descriptor in the set of ~$3\times10^9$ potential descriptors, we cannot be certain that a subspace size of 100,000 is sufficient to find the best descriptor. However, the identification of $\tau$ within a subspace as small as 15,000 suggests that a subspace size of 100,000 is sufficiently large to efficiently screen the much larger descriptor space. We have also conducted a test on this primary feature space ($\Phi_0 = [n_A, n_B, n_X, r_A, r_B, r_X, r_A/r_B, r_A/r_X, r_B/r_X]$) with a subspace size of 500,000. Even after increasing the subspace size by 5×, $\tau$ remains the highest performing descriptor (classification accuracy of 92% on the 576 compound set). An important distinction between the SISSO approach described here and by Ouyang *et al.* in (*32*) is the choice of sparsifying operator (SO). In this work, domain overlap is used to *rank* the features in SISSO, but a decision tree with max depth of two is used as the SO (instead of domain overlap) to *identify* the best descriptor of those selected by SISSO. This alternative SO is used to decrease the leverage of individual data points as the experimental labeling of perovskite/nonperovskite is prone to some ambiguity based on synthesis conditions, defects, and other experimental considerations.

The benefit of including the radii ratios in $\Phi_0$ is made clear by comparing the performance of $\tau$ to the best descriptor obtained using the minimal primary feature space with $\Phi_0 = [n_A, n_B, n_X, r_A, r_B, r_X]$. Repeating the procedure used to identify $\tau$ yields a $\Phi_3$ with ~$1\times10^8$ potential descriptors. The best 1D descriptor was found to be $\frac{r_B}{n_X(r_A-r_B)} + \frac{r_B}{r_A} - \frac{r_X}{r_B}$ with classification accuracy of 89%.

**Alternative features**

We also consider the effects of including properties outside of those required to compute $t$ or $\tau$. Beginning with $\Phi_0 = [n_A, n_B, n_X, r_A, r_B, r_X, r_{cov,A}, r_{cov,B}, r_{cov,X}, IE_A, IE_B, IE_X, \chi_A, \chi_B, \chi_X]$, where $r_{cov,i}$ is the empirical covalent radius of neutral element $i$, $IE_i$ is the empirical first ionization energy of neutral element $i$, and $\chi_i$ is the Pauling electronegativity of element $i$, all taken from WebElements,(*45*) an aggregation of a number of references which are available within. Repeating the procedure used to identify $\tau$ results in ~$6\times10^{10}$ potential descriptors in $\Phi_3$. The best performing 1D descriptor was found to be $\frac{r_A/r_B - \sqrt{\chi_X}}{r_{cov,X}/r_B - r_{cov,A}/r_{cov,X}}$ with classification accuracy of 90%, lower than $\tau$ which makes use of only the oxidation states and ionic radii, and only slightly higher than the accuracy of the descriptor obtained using the minimal feature set.

**Increasing dimensionality**

To assess the performance of descriptors with increased dimensionality, following the approach to higher dimensional descriptor identification using SISSO described in (*32*), the residuals from classification by $\tau$ (those misclassified by the decision tree, **Fig. 2b**) are used as the target property in the search for a second dimension to include with $\tau$. From the same set of ~$3\times10^9$ potential descriptors constructed to identify $\tau$, the 100,000 1D descriptors that best classify the 41 training set compounds misclassified by $\tau$ are identified based on domain overlap. Each of these 100,000 descriptors are paired with $\tau$ and the performance of each 2D descriptor was assessed using a decision tree with max depth of two. The best performing 2D descriptor was found to be $\left(\tau, \frac{|r_A r_X/r_B^2 - n_B r_A/r_B|}{|r_A r_B/r_X^2 - r_A/r_B + n_B|}\right)$ with a classification accuracy of 95% on the 576 compound set. Improvements are expected to diminish as the dimensionality increases further due to the iterative nature of SISSO and the higher order residuals used for subspace selection. Although the second dimension leads to slightly improved classification performance on the experimental set compared with $\tau$, the simplicity and monotonicity of $\tau$, which enables physical interpretation and the extraction of meaningful probabilities, support its selection instead of the more complex 2D descriptor. The benefits and capabilities of having a meaningfully probabilistic one-dimensional tolerance factor, such as $\tau$, are described in detail within the main text.

**Potential for overfitting**

The SISSO algorithm as implemented here *selects* $\tau$ from a space of ~3×10$^9$ candidate descriptors and the only parameter that is *fit* is the optimum value of $\tau$ that defines the decision boundary for classification as perovskite or nonperovskite, $\tau = 4.18$. This decision boundary is optimized using a decision tree to maximize the classification accuracy on the training set of 460 compounds. In this case, Gini impurity is minimized to optimize the decision boundary, but alternative cost functions based on Kullback-Leibler divergence or classification accuracy (e.g., $\ell_2$) would find the same decision boundary. The SISSO descriptor identification is done from billions of candidates, but these functions comprise a discrete set, i.e., they form a basis in a large dimensional space where the number of training points is the dimensionality of the space, which is not densely covered by the basis functions. Therefore, the selection of only one function, $\tau$, cannot overfit the data. However, if some physical mechanism determining the stability of perovskites is not represented in the training set, it might be missed by the learned formula (here, $\tau$) and therefore the generalizability of the model would be hampered. However, the 94% accuracy achieved by $\tau$ on the excluded set of 116 compounds shows that $\tau$ *can* generalize outside of the training data.

**Alternative radii for more covalent compounds**

Ionic radii are required inputs for $\tau$ (and $t$) and although the Shannon effective ionic radii are ubiquitous in solid state materials research, a new set of $B^{2+}$ radii were recently proposed for 18 cations to account for how their effective cationic radii vary as a function of increased covalency with the heavier halides.(*18*) These revised radii apply to 129 of the 576 experimentally characterized compounds compiled in this dataset (62% of halides). Employing these revised radii results in a 5% decrease in the accuracy of $\tau$ to 86% for these 129 compounds compared to a classification accuracy of 91% using the Shannon radii for these same compounds. The application of $\tau$ using Shannon radii for presumably covalent compounds is further validated by noting that $\tau$ correctly classifies 37/40 compounds which contain Sn or Pb and achieves an accuracy of 91% for 141 compounds with $X = $ Cl$^-$, Br$^-$, or I$^-$. In addition to the higher accuracy achieved by $\tau$ when using Shannon radii, we note that the Shannon radii are more comprehensive than these revised radii in (*18*), applying to more ions, oxidation states, and coordination environments and are thus recommended for the calculation of $\tau$.

**Computer packages used**

SISSO was performed using *Fortran 90*. Platt's scaling(*33*) was used to extract classification probabilities for $\tau$ by fitting a logistic regression model on the decision tree classifications using 3-fold cross-validation. Decision tree fitting and Platt scaling were performed within the *Python* package, *scikit-learn*. Data visualizations were generated within the *Python* packages *matplotlib* and *seaborn*.

**List of Supplementary Materials**

**Fig. S1.** Classification accuracy of $t$ and $\tau$ as a function of the anion, *X*.

**Fig. S2.** $P(\tau)$ vs. $\tau$.

**Fig. S3.** Structure map for classification using the 2D descriptor $(t, \mu)$.

**Table S1.** 576 $ABX_3$ compounds with experimental characterization used for identifying and testing $\tau$.

**Table S2.** Confusion matrices for classification with $t$ and $\tau$.

**Table S3.** DFT-calculated data associated with Fig. 2D.

**Table S4.** Double perovskite oxide and halide data from the ICSD and new predictions.

**References**

1. L. Pauling, The principles determining the structure of complex ionic crystals. *J. Am. Chem. Soc.* **51**, 1010-1026 (1929).
2. S. M. Woodley, R. Catlow, Crystal structure prediction from first principles. *Nat. Mater.* **7**, 937 (2008).
3. S. Kirkpatrick, C. D. Gelatt, M. P. Vecchi, Optimization by simulated annealing. *Science* **220**, 671-680 (1983).
4. J. P. K. Doye, D. J. Wales, Thermodynamics of global optimization. *Phys. Rev. Lett.* **80**, 1357-1360 (1998).
5. S. Goedecker, Minima hopping: An efficient search method for the global minimum of the potential energy surface of complex molecular systems. *J. Chem. Phys.* **120**, 9911-9917 (2004).
6. A. R. Oganov, A. O. Lyakhov, M. Valle, How evolutionary crystal structure prediction works—and why. *Acc. Chem. Res.* **44**, 227-237 (2011).
7. S. Curtarolo *et al.*, The high-throughput highway to computational materials design. *Nat. Mater.* **12**, 191 (2013).
8. L. M. Ghiringhelli, J. Vybiral, S. V. Levchenko, C. Draxl, M. Scheffler, Big data of materials science: Critical role of the descriptor. *Phys. Rev. Lett.* **114**, 105503 (2015).
9. V. M. Goldschmidt, Die Gesetze der Krystallochemie. *Naturwissenschaften* **14**, 477-485 (1926).
10. J. Hwang *et al.*, Perovskites in catalysis and electrocatalysis. *Science* **358**, 751-756 (2017).
11. C. Duan *et al.*, Readily processed protonic ceramic fuel cells with high performance at low temperatures. *Science*, (2015).
12. R. E. Cohen, Origin of ferroelectricity in perovskite oxides. *Nature* **358**, 136 (1992).
13. T. Yi *et al.*, Investigating the intercalation chemistry of alkali ions in fluoride perovskites. *Chem. Mater.* **29**, 1561-1568 (2017).
14. J.-P. Correa-Baena *et al.*, Promises and challenges of perovskite solar cells. *Science* **358**, 739-744 (2017).
15. M. V. Kovalenko, L. Protesescu, M. I. Bodnarchuk, Properties and potential optoelectronic applications of lead halide perovskite nanocrystals. *Science* **358**, 745-750 (2017).
16. H. Zhang, N. Li, K. Li, D. Xue, Structural stability and formability of $ABO_3$-type perovskite compounds. *Acta Crystallogr. Sect. B* **63**, 812-818 (2007).
17. C. Li *et al.*, Formability of $ABX_3$ (X = F, Cl, Br, I) halide perovskites. *Acta Crystallogr. Sect. B* **64**, 702-707 (2008).
18. W. Travis, E. N. K. Glover, H. Bronstein, D. O. Scanlon, R. G. Palgrave, On the application of the tolerance factor to inorganic and hybrid halide perovskites: A revised system. *Chem. Sci.* **7**, 4548-4556 (2016).
19. W. Li *et al.*, Chemically diverse and multifunctional hybrid organic–inorganic perovskites. *Nat. Rev. Mater.* **2**, 16099 (2017).
20. R. Shannon, Revised effective ionic radii and systematic studies of interatomic distances in halides and chalcogenides. *Acta Crystallographica Section A* **32**, 751-767 (1976).
21. M. W. Lufaso, P. M. Woodward, Prediction of the crystal structures of perovskites using the software program SPuDS. *Acta Crystallogr. Sect. B* **57**, 725-738 (2001).

22. G. Kieslich, S. Sun, A. K. Cheetham, Solid-state principles applied to organic-inorganic perovskites: new tricks for an old dog. *Chemical Science* **5**, 4712-4715 (2014).
23. C. Li, K. C. K. Soh, P. Wu, Formability of $ABO_3$ perovskites. *J. Alloys Compd.* **372**, 40-48 (2004).
24. M. Becker, T. Kluner, M. Wark, Formation of hybrid $ABX_3$ perovskite compounds for solar cell application: first-principles calculations of effective ionic radii and determination of tolerance factors. *Dalton Transactions* **46**, 3500-3509 (2017).
25. G. Pilania, P. V. Balachandran, J. E. Gubernatis, T. Lookman, Classification of $ABO_3$ perovskite solids: a machine learning study. *Acta Crystallogr. Sect. B* **71**, 507-513 (2015).
26. G. Pilania, P. V. Balachandran, C. Kim, T. Lookman, Finding new perovskite halides via machine learning. *Front. Mater.* **3**, 19 (2016).
27. P. V. Balachandran *et al.*, Predictions of new $ABO_3$ perovskite compounds by combining machine learning and density functional theory. *Physical Review Materials* **2**, 043802 (2018).
28. X.-G. Zhao *et al.*, Cu–In Halide Perovskite Solar Absorbers. *Journal of the American Chemical Society* **139**, 6718-6725 (2017).
29. Q. Sun, W.-J. Yin, Thermodynamic Stability Trend of Cubic Perovskites. *Journal of the American Chemical Society* **139**, 14905-14908 (2017).
30. D. M. Helen, Crystal structure of double oxides of the perovskite type. *Proc. Phys. Soc.* **58**, 133 (1946).
31. A. Feteira, D. C. Sinclair, K. Z. Rajab, M. T. Lanagan, Crystal structure and microwave dielectric properties of alkaline-earth hafnates, $AHfO_3$ (A=Ba, Sr, Ca). *J. Am. Ceram. Soc.* **91**, 893-901 (2008).
32. R. Ouyang, S. Curtarolo, E. Ahmetcik, M. Scheffler, L. M. Ghiringhelli, SISSO: A compressed-sensing method for identifying the best low-dimensional descriptor in an immensity of offered candidates. *Physical Review Materials* **2**, 083802 (2018).
33. J. Platt, Probabilistic outputs for support vector machines and comparisons to regularized likelihood methods. *Advances in large margin classifiers* **10**, 61-74 (1999).
34. M. R. Filip, F. Giustino, The geometric blueprint of perovskites. *Proceedings of the National Academy of Sciences* **115**, 5397-5402 (2018).
35. J. P. Perdew, K. Burke, M. Ernzerhof, Generalized Gradient Approximation Made Simple. *Physical Review Letters* **77**, 3865-3868 (1996).
36. J. Schmidt *et al.*, Predicting the Thermodynamic Stability of Solids Combining Density Functional Theory and Machine Learning. *Chemistry of Materials* **29**, 5090-5103 (2017).
37. F. A. Faber, A. Lindmaa, O. A. von Lilienfeld, R. Armiento, Machine Learning Energies of 2 Million Elpasolite $ABC_2D_6$ Crystals. *Physical Review Letters* **117**, 135502 (2016).
38. T. Xie, J. C. Grossman, Crystal Graph Convolutional Neural Networks for an Accurate and Interpretable Prediction of Material Properties. *Physical Review Letters* **120**, 145301 (2018).
39. S. Kirklin *et al.*, The Open Quantum Materials Database (OQMD): assessing the accuracy of DFT formation energies. *npj Computational Materials* **1**, 15010 (2015).
40. P. V. Kamat, J. Bisquert, J. Buriak, Lead-Free Perovskite Solar Cells. *ACS Energy Letters* **2**, 904-905 (2017).
41. M. Hellenbrandt, The Inorganic Crystal Structure Database (ICSD)—Present and Future. *Crystallography Reviews* **10**, 17-22 (2004).
42. W. Li, E. Ionescu, R. Riedel, A. Gurlo, Can we predict the formability of perovskite oxynitrides from tolerance and octahedral factors? *Journal of Materials Chemistry A* **1**, 12239-12245 (2013).
43. E. T. McClure, M. R. Ball, W. Windl, P. M. Woodward, $Cs_2AgBiX_6$ (X = Br, Cl): New Visible Light Absorbing, Lead-Free Halide Perovskite Semiconductors. *Chemistry of Materials* **28**, 1348-1354 (2016).
44. S. O. Choi, M. Penninger, C. H. Kim, W. F. Schneider, L. T. Thompson, Experimental and Computational Investigation of Effect of Sr on NO Oxidation and Oxygen Exchange for $La_{1-x}Sr_xCoO_3$ Perovskite Catalysts. *ACS Catalysis* **3**, 2719-2728 (2013).
45. WebElements.com.

**Acknowledgments: General:** The authors thank Aaron Holder for helpful discussions regarding the manuscript. **Funding:** The project received funding from the European Union's Horizon 2020 research and innovation program under grant agreement no. 676580 with The Novel Materials Discovery (NOMAD) Laboratory, a European Center of Excellence. C.J.B. acknowledges support from a US Department of Education Graduate Assistantship in Areas of National Need. C.S. gratefully acknowledges funding by the Alexander von Humboldt Foundation. C.B.M acknowledges support from NSF award CBET-1433521, which was cosponsored by the NSF, and the U.S. DOE, EERE, Fuel Cell Technologies Office and from DOE award EERE DE-EE0008088. Part of this research was performed using computational resources sponsored by the U.S. Department of Energy's Office of Energy Efficiency and Renewable Energy and located at the National Renewable Energy Laboratory. **Author contributions:** MS and CJB conceived the idea. CJB, CS, and BRG designed the studies. CJB performed the studies. CJB, CS, and BRG analyzed the results and wrote the manuscript. RO provided the SISSO algorithm and facilitated its implementation. CBM, LMG, and MS supervised the project. All the authors discussed the results and implications and edited the manuscript. **Competing interests:** The authors declare no competing financial interests. **Data availability:** A repository containing all files necessary for classifying $ABX_3$ and $AA'BB'(XX')_6$ compositions as perovskite or nonperovskite using $\tau$ is available at https://github.com/CJBartel/perovskite-stability. A graphical interface allowing users to classify compounds with $\tau$ is also available at https://analytics-toolkit.nomad-coe.eu. The classification of all compounds shown in the manuscript is available in the Supplementary Materials.

## Supplementary Materials

**Table S1. 576 $ABX_3$ used for training and testing $\tau$ (see TableS1_exp_data.csv).**
Within this table – *exp_label* = 1 corresponds with experimentally labeled perovskites and *exp_label* = −1, nonperovskites; *is_train* = 1 corresponds with a training set compound and *is_train* = −1 a test set compound; $\tau$ is provided as *tau*, classification using $\tau$ as *tau_pred*, classification using *t* as *t_pred*, and $\tau$-derived probabilities, P($\tau$) as *tau_prob*. *A, B, X, $n_A$, $n_B$, $n_X$, $r_A$, $r_B$, $r_X$,* and *t* are also provided and named as they are in the text.

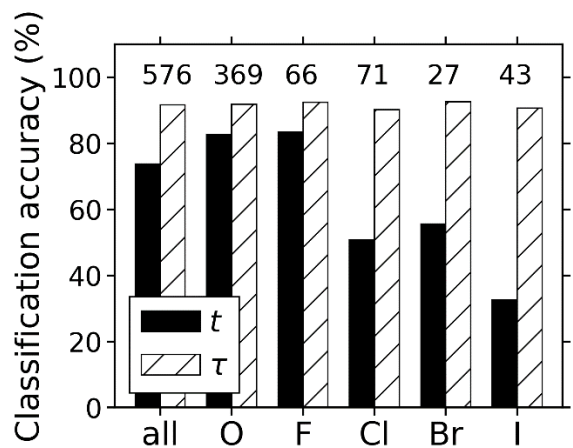

**Figure S1. Comparing the performance of *t* and $\tau$ by composition.**
Classification accuracy on the full set of 576 experimentally characterized $ABX_3$ solids (*all*) and by compounds containing $X$ = $O^-$, $F^-$, $Cl^-$, $Br^-$, $I^-$. The number appearing above each pair of columns corresponds with the number of compounds evaluated within each set.

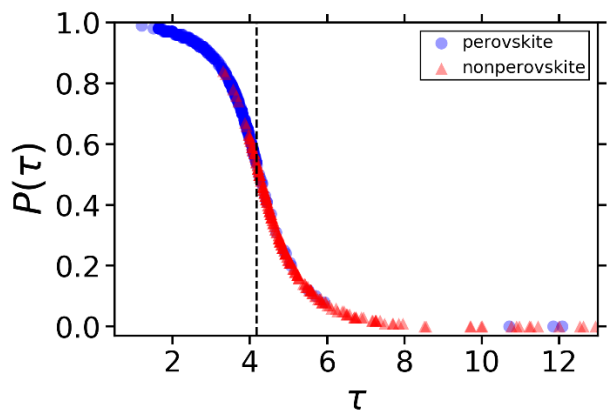

**Figure S2. Sigmoidal relationship between $P(\tau)$ and $\tau$.**
A comparison of Platt-scaled classification probabilities, $P(\tau)$, versus $\tau$ for the 576 experimentally characterized $ABX_3$ solids. The outlying compounds at $\tau > 10$ that are labeled perovskite yet have small $P(\tau)$ are $PuVO_3$, $AmVO_3$, and $PuCrO_3$, which may indicate poorly defined radii or experimental characterization.

**Table S2. Confusion matrices for $\tau$ (above) and *t* (below).**
N = number of samples; TP = true positives (experimentally realized perovskites which are predicted perovskite); FP = false positives (predicted perovskites which are not experimentally realized perovskites); TN = true negatives (predicted nonperovskites which are experimentally realized nonperovskites); FN = false negatives (predicted nonperovskites which are experimentally realized perovskites); precision = TP/(TP+FP); recall = TP/(TP+FN); F1 score = (2*precision*recall)/(precision + recall); kappa = Cohen's Kappa score; accuracy = (TP+TN)/N.

```
tau:
+-------+-----+-----+----+-----+----+-----------+--------+----------+-------+----------+
| anion |  N  | TP  | FP | TN  | FN | precision | recall | F1 score | kappa | accuracy |
+-------+-----+-----+----+-----+----+-----------+--------+----------+-------+----------+
|  all  | 576 | 293 | 28 | 235 | 20 |   0.913   | 0.936  |  0.924   | 0.832 |  0.917   |
|   O   | 369 | 217 | 15 | 122 | 15 |   0.935   | 0.935  |  0.935   | 0.826 |  0.919   |
|   F   |  66 |  50 |  4 |  11 |  1 |   0.926   | 0.980  |  0.952   | 0.768 |  0.924   |
|  Cl   |  71 |  18 |  7 |  46 |  0 |   0.720   | 1.000  |  0.837   | 0.769 |  0.901   |
|  Br   |  27 |   5 |  2 |  20 |  0 |   0.714   | 1.000  |  0.833   | 0.787 |  0.926   |
|   I   |  43 |   3 |  0 |  36 |  4 |   1.000   | 0.429  |  0.600   | 0.557 |  0.907   |
+-------+-----+-----+----+-----+----+-----------+--------+----------+-------+----------+
t:
+-------+-----+-----+-----+-----+----+-----------+--------+----------+-------+----------+
| anion |  N  | TP  | FP  | TN  | FN | precision | recall | F1 score | kappa | accuracy |
+-------+-----+-----+-----+-----+----+-----------+--------+----------+-------+----------+
|  all  | 576 | 295 | 133 | 130 | 18 |   0.689   | 0.942  |  0.796   | 0.453 |  0.738   |
|   O   | 369 | 216 |  48 |  89 | 16 |   0.818   | 0.931  |  0.871   | 0.610 |  0.827   |
|   F   |  66 |  49 |   9 |   6 |  2 |   0.845   | 0.961  |  0.899   | 0.432 |  0.833   |
|  Cl   |  71 |  18 |  35 |  18 |  0 |   0.340   | 1.000  |  0.507   | 0.207 |  0.507   |
|  Br   |  27 |   5 |  12 |  10 |  0 |   0.294   | 1.000  |  0.455   | 0.236 |  0.556   |
|   I   |  43 |   7 |  29 |   7 |  0 |   0.194   | 1.000  |  0.326   | 0.073 |  0.326   |
+-------+-----+-----+-----+-----+----+-----------+--------+----------+-------+----------+
```

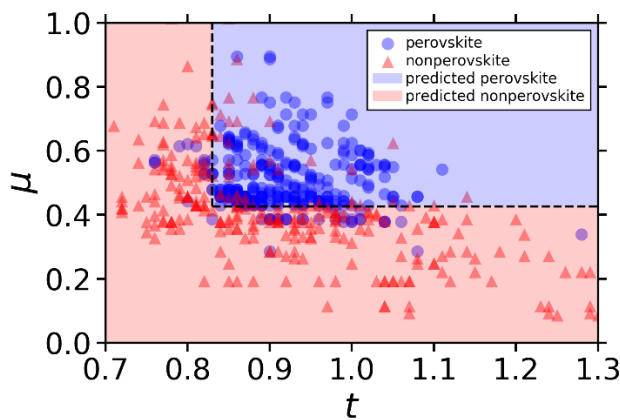

**Figure S3. ($t, \mu$) structure map for 576 $ABX_3$ solids.**
A decision tree trained on all compounds finds the optimal bounds for perovskites stability to occur for $t > 0.830$ and $\mu > 0.426$, which results in a classification accuracy of 85%.

**Table S3. Additional information associated with Fig. 2D (see Table S3_dft_data.csv).**
Columns are named as described for Table S1 with the additional columns: *source* – corresponding with the DOI from which the DFT decomposition enthalpy was obtained and *dHdec (meV/atom)* – the PBE-computed decomposition enthalpy of the cubic structure.

**Table S4. Double perovskite oxides and halides (see TableS4_A2BBX6_data.csv).**
All charge-balanced compounds described in the context of **Fig. 3** are provided. Columns are named as described for **Table S1** and **Table S3**. An additional column, *icsd_label*, provides the assignment of perovskite/nonperovskite extracted from the ICSD if available.